\documentclass[prb,preprintnumbers,amsfonts,amssymb,amsmath,floats,twocolumn,aps]{revtex4-2}

\usepackage[pdftex]{graphicx}
\usepackage[caption=false]{subfig}
\usepackage{dcolumn}
\usepackage{bm}
\usepackage{color}
\usepackage{booktabs}

\usepackage[pdftex,colorlinks=true,linkcolor=blue,citecolor=blue,filecolor=blue]{hyperref}

\newcommand{\bi}{\begin{itemize}}
\newcommand{\ei}{\end{itemize}}
\newcommand{\be}{\begin{equation}}
\newcommand{\ee}{\end{equation}}
\newcommand{\ba}{\begin{eqnarray}}
\newcommand{\ea}{\end{eqnarray}}

\begin{document} 

\title{Competing States in the Two-Dimensional Frustrated Kondo-Necklace Model}
\author{Matthias Peschke$^{1,2}$, Boris Ponsioen$^1$ and Philippe Corboz$^1$}
\affiliation{{\bf 1} Institute for Theoretical Physics and Delta Institute for Theoretical Physics, University of Amsterdam, Science Park 904,
1098 XH Amsterdam, The Netherlands}
\affiliation{{\bf 2} Department of Physics, University of Hamburg, Jungiusstra{\ss}e 9, D-20355 Hamburg, Germany}
\begin{abstract}
  The interplay between Kondo screening, indirect magnetic interaction and geometrical frustration is studied in the two-dimensional Kondo-necklace model on the triangular lattice.
  Using infinite projected entangled pair states (iPEPS),
  we compute the ground state as a function of the antiferromagnetic local Kondo interaction $J_K$ and the Ising-type direct spin-spin interaction $I_z$.
  As opposed to previous studies, we do not find partial Kondo screening (PKS) in the isotropic limit $I_z=0$ but the same behavior as in the unfrustrated case, i.e.
  a direct phase transition between the paradigmatic phases of the Doniach competition:
  (i) a disordered phase consisting of local spin-singlets at strong $J_K$ and (ii) a magnetically ordered phase at weak $J_K$.
  For $I_z>0$, we find a PKS ground state but again in opposite to previous studies,
  we find that the PKS ground state is in strong competition with a second ground state candidate not found before.
  This state is characterized by a strongly polarized central spin in each hexagon and its anti-parallel, weakly polarized (i.e. partially screened) neighbors.
  We name it central spin (CS) phase.
\end{abstract} 

\maketitle 

\section{Introduction}
\label{sec:intro}

Two-dimensional strongly interacting lattice models constitute one of the most
challenging problems in condensed matter theory.
The complexity is especially high in systems with frustration in the presence of large quantum fluctuations based on the low dimensionality.
For example, the
unconventional $d$-wave superconductivity in cuprates arises when introducing frustration by doping the antiferromagnetic Mott insulator~\cite{BM86, WAT87, LNW06, KKN15, RJR19, QSA22}.
Heavy-fermion compounds constitute another example for exotic quantum phases driven by
the famous Doniach competition between a nonmagnetic Kondo screened and a RKKY induced magnetic ground state~\cite{Don77}.
Here, additional frustration can enter the competition via a lattice structure which is incompatible with the magnetic order.
The compounds CePdAl and $\text{UNi}_4\text{B}$ are layered materials which form a triangular lattice which induces geometrical frustration of the antiferromagnetic order~\cite{OMN08,FLH17,MDN94,OKF07}.
In both materials, experiments suggests an exotic ground state which exhibits partial Kondo screening (PKS)~\cite{OMN08,FLH17,MDN94,OKF07}, i.e.,
a trade-off for the geometrical frustration in which one-third of the moments are screened while the remnant two-third form a magnetically ordered state.

The quantum Monte Carlo approach is a powerful tool to study these interacting two-dimensional systems
but it has weaknesses for the cases where the exotic phases are driven by frustration, due to the infamous sign problem~\cite{troyer2005}.
On the other hand, within the last decade,
great progress was achieved in the theoretical description of these systems by using tensor network related approaches~\cite{Verstraete08,Sch11,EV11,orus14,CPS21}.
For two-dimensional systems, projected-entangled pair states (PEPS)~\cite{VC04,Nishio2004,Murg2007} have proven to offer a competitive tool for studying ground state properties.
In particular the infinite PEPS (iPEPS) algorithm~\cite{JOV08,COB10}, which is working directly in the thermodynamic limit, was used for several strongly frustrated two-dimensional systems, see e.g. Refs.~\cite{corboz14_shastry, liao16, niesen17, chen18, jahromi18, lee18, kshetrimayum19b,boos19, haghshenas19,lee20, hasik21, liu21}.

Several theoretical investigations of the influence of geometrical frustration on the Doniach competition in heavy-fermion models have proposed
that PKS can release the geometrical frustration.
For the two-dimensional Anderson lattice, Hartree-Fock studies~\cite{HUM11,HUM12} suggest a PKS ground state at and away from half-filling.
The PKS ground state, however, was not found when using dynamical mean-field theory at half-filling but only at finite doping~\cite{AAP15}.
For the two-dimensional Kondo lattice, a PKS ground state was found by applying variational Monte Carlo (VMC)~\cite{MNY10} but for a one dimensional zig-zag chain,
DMRG simulations discovered a spontaneous dimerization as a different trade-off for the geometrical frustration~\cite{PRP18,PWP19}.
Finally, VMC simulations also predict PKS for the two-dimensional Kondo-necklace model (KNM) on the triangular and Kagom\'{e} lattice~\cite{MNY10,MNY11}.
Apart from the DMRG simulations for the one-dimensional case,
all the mentioned studies depend on approximations which might not be able to describe the highly complicated physics emerging from the geometrical frustration.
Additionally, the VMC simulations were only performed on finite clusters of total size up to $N=24$ which might not be enough to describe truly two-dimensional behavior.

The purpose of the present study is the systematic investigation of the influence of geometrical frustration on the Doniach competition in the 2D Kondo necklace model on the triangular lattice
by employing the iPEPS framework.
Our numerical simulations demonstrate the absence of any intermediate phase between the two characteristic phases of the Doniach competition (disordered and magnetic) in the isotropic limit
for the triangular lattice.
For the anisotropic case however, a new possibility for releasing the geometrical frustration is discovered which is found to be in strong competition with the previously found PKS ground state.
It is characterized by a strongly polarized central spin in each hexagon and its anti-parallel weakly polarized neighbors.
We name it \emph{central spin} (CS) phase.

\section{Model}
\label{sec:mod}

We consider the Kondo-necklace model, i.e., two layers of quantum spins with $S=1/2$ coupled via a local antiferromagnetic Kondo exchange interaction with coupling constant $J_K>0$.
In both layers, the spins are positioned on the vertices of a two-dimensional triangular lattice.
The upper layer constitutes an isotropic triangular antiferromagnetic Heisenberg model with coupling constant $J>0$ while the lower layer forms a triangular Ising model with coupling constant $I_z\geq 0$.  
The Hamiltonian reads
\begin{equation}
{H} = J \sum_{\langle i,j \rangle} \pmb{s}_{i} \pmb{s}_{j} + J_K \sum_{i} \pmb{S}_{i}\pmb{s}_{i} + I_z \sum_{\langle i,j \rangle} S^z_{i}S^z_{j}.
\label{eq:ham}
\end{equation}

\begin{figure}
\subfloat[Bilayer triangular lattice \label{sfig:model}]{%
  \includegraphics[width=0.49\linewidth]{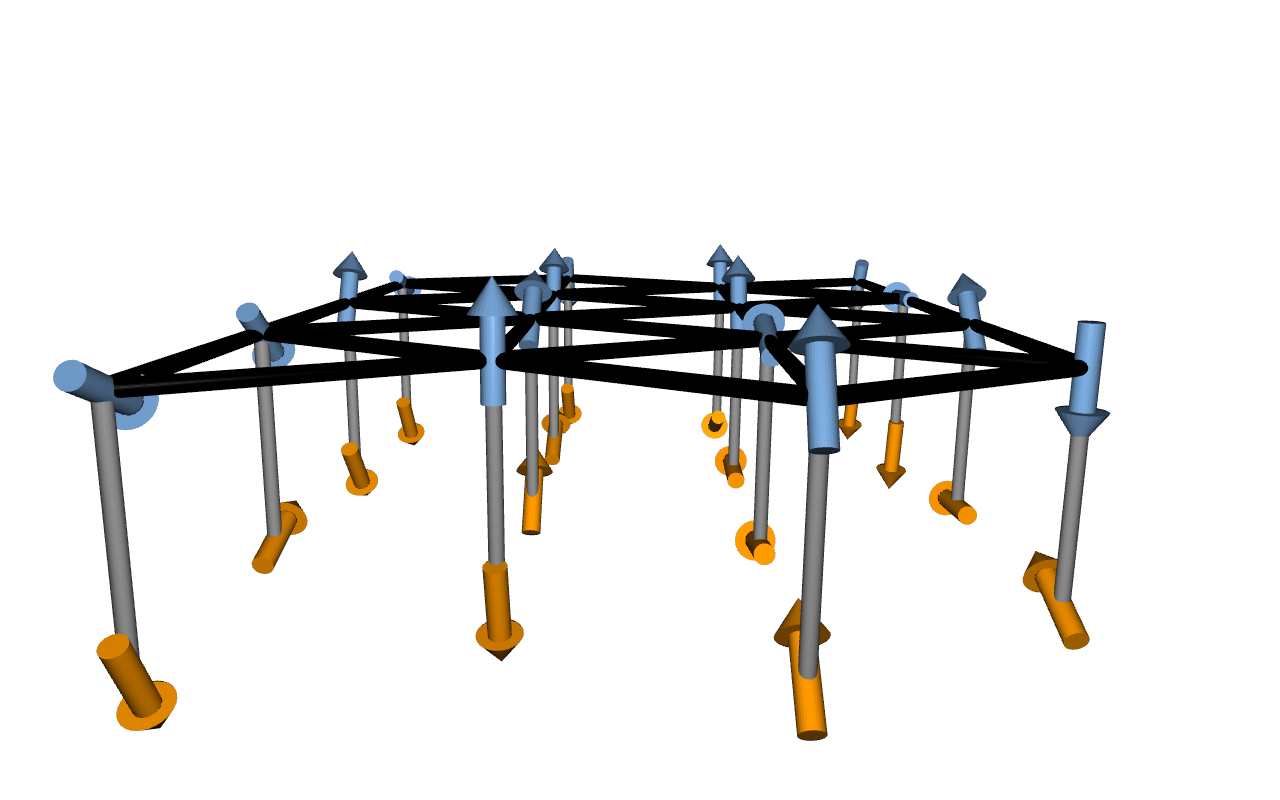}%
}
\subfloat[iPEPS unit cell \label{sfig:ipeps_unitcell}]{%
  \includegraphics[width=0.49\linewidth]{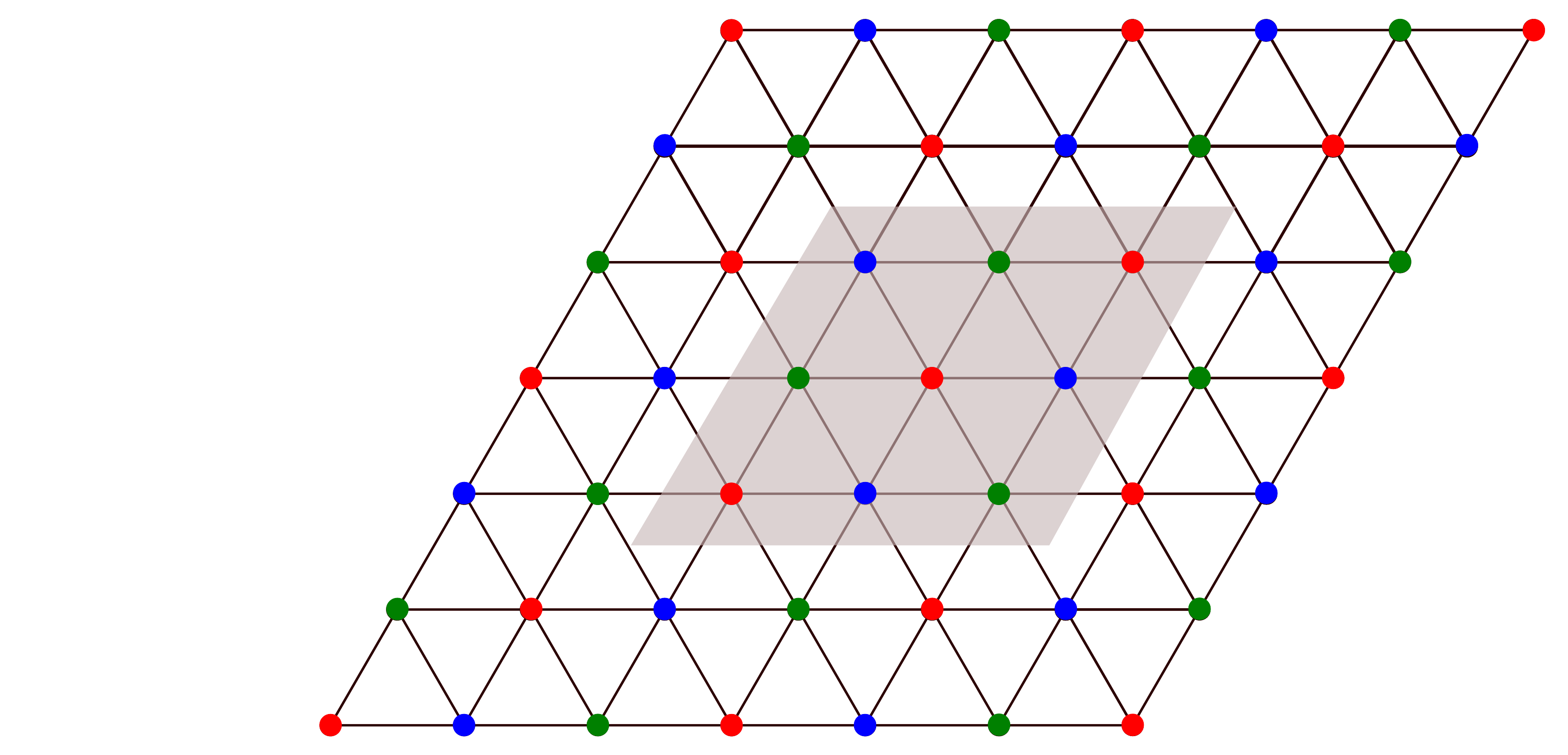}%
} \hfill
\caption{Left: sketch of the Kondo-necklace model with vanishing $I_z$.
  In this case, the lower layer (orange arrows) has no direct interaction but is only coupled indirectly via the upper layer (blue arrows).
  Right: the $3\times 3$ unit cell (shaded gray) used in our iPEPS simulations.
  Additionally, we enforce a pattern so that the unit cell contains three independent tensors (colored red, green and blue)
  which enables a breaking of the translational symmetry into three sublattices.}
\label{fig:sketches}
\end{figure}

Here, $\pmb{s}_i$ ($\pmb{S}_i$) denotes the spin operator in the upper (lower) layer and the sum in
the first and third term is taken over all adjacent sites $i=(i_x,i_y)$ and $j=(j_x,j_y)$ in the triangular lattice.
We show a sketch of the geometry for vanishing $I_z$, i.e. the isotropic case,  in Fig.~\ref{sfig:model}.

This model can describe the qualitative physics of the full Kondo lattice model at half-filling as long as the charge degrees of freedom of the conduction electron layer are frozen.
This can arise by a large Hubbard interaction within the electron layer or by dominant Kondo screening.

\section{Methods}
\label{sec:methods}
To compute the ground state of the Hamiltonian from Eq.~\ref{eq:ham},
we employ the infinite projected-entangled pair state (iPEPS)~\cite{VC04, JOV08, PBT15} approach directly in the thermodynamic limit.

In the iPEPS framework the ground state wave function $|\Psi\rangle$ is parameterized by $N_c$ rank-5 tensors $A^\sigma_{udlr}(i)$
where $i=(i_x,i_y)$ denotes a site in the unit cell with $N_c=L_x\cdot L_y$ sites.
Such a parameterization obeys the ``area law'' for the entanglement scaling of ground states of short range two-dimensional Hamiltonians~\cite{ECP10}
and therefore represents an efficient variational ansatz for such states.
The elementary site tensors $A(i)$ which build the iPEPS have one physical index $\sigma$ and four auxiliary indices of bond dimension $D$ which determines the accuracy of the ansatz.

In order to find an iPEPS which faithfully represents the ground state, one needs to optimize the variational parameters in the $A$-tensors.
This can be achieved by an imaginary time evolution or by gradient based optimization methods.
When evolving the state in imaginary time, the bond dimension $D$ grows so that a renormalization step is necessary.
A common renormalization method is the simple update~\cite{JWX08} which performs the renormalization by using an approximated environment of the network.
More accurate renormalization can be achieved with the cluster~\cite{WV11,LCB14} and full update~\cite{JOV08,PBT15}.
For the latter, the environment of the infinite 2D double layer network needs to be computed to perform the renormalization.
This is performed with the corner transfer matrix (CTM)~\cite{NO96, OV09, CRT14} approach which introduces an effective virtual bond dimension $\chi$ in the environment.
We find that the simple update is not able to capture the relevant physics
and used the cluster update instead for finding initial guesses for the gradient based optimization.
Here, we use the variational method described in Ref.~\onlinecite{Cor16} and a limited memory variant of the Broyden–Fletcher–Goldfarb–Shanno algorithm (L-BFGS) based on automatic differentiation~\cite{LLW19, PFC22} (AD).
We employ the U(1) symmetry of the Hamiltonian to reduce the computational cost~\cite{SPR11, BCO11} for some of the simulations.

\section{Results}
\label{sec:res}

\begin{figure}
  \includegraphics{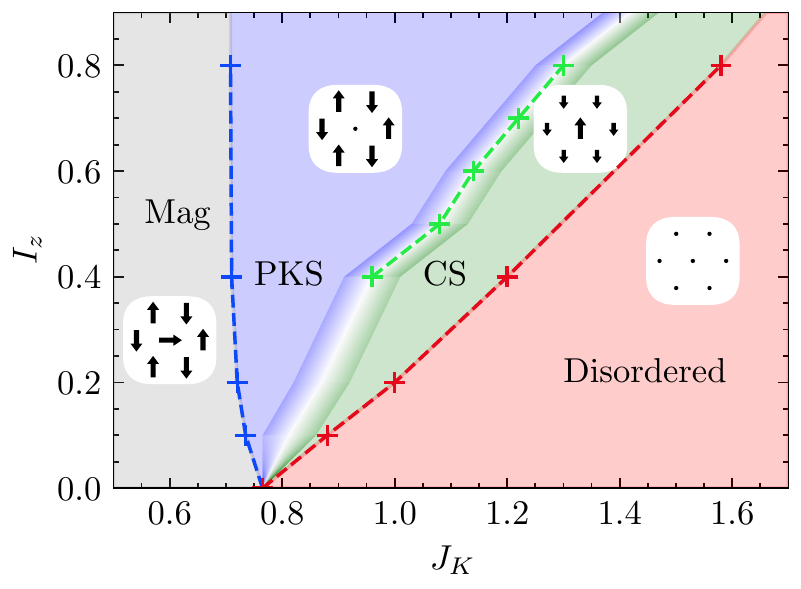}
\caption{
  $J_K$--$I_z$ phase diagram based on iPEPS simulations with bond dimension $D=4$.
  Gray and red denote the characteristic phases of the Doniach competition: (i) disordered (red) at strong $J_K$ and magnetic (gray) at weak $J_K$.
  The geometrical frustration leads to an intermediate region in the phase diagram (blue and green) in which two ground state candidates are found
  which are in strong competition over the entire intermediate region:
  (i) partial Kondo screening (blue) is realized at weaker $J_K$ and (ii) central spin phase (green) is realized at stronger $J_K$.
  The white region indicates a broad transition regime (possibly a mixed phase) between PKS and CS.
  Insets: magnetization patterns of the lower layer in the respective phases.
}
\label{fig:sketch_pd}
\end{figure}

Figure~\ref{fig:sketch_pd} represents the $J_K$--$I_z$ phase diagram which gives a broad overview of the different phases.
As expected for heavy-fermion models, we find a disordered phase at strong $J_K$ which is colored red in Fig.~\ref{fig:sketch_pd}
and a magnetic phase at weak $J_K$ which is colored gray. The magnetic phase has clock order for finite anisotropy $I_z$ (see inset)
which turns smoothly into a $120^\circ$ phase for $I_z\to 0$.
In between these two phases, the geometrical frustration leads to the emergence of an exotic ground state.
However, we find that \emph{two} candidates are in strong competition.
The \emph{first} one is the PKS ground state (blue in Fig.~\ref{fig:sketch_pd}) which was proposed as the ground state of the KNM by a previous VMC study~\cite{MNY10}.
The \emph{second} one is the CS ground state (green in Fig.~\ref{fig:sketch_pd}) which was not found by any previous work.
It is characterized by a strongly polarized central spin in each hexagon which is surrounded by weakly polarized anti-parallel adjacent spins.

The different phases are all characterized by symmetry breaking. The disordered phase preserves all symmetries of the Hamiltonian,
i.e. the translational symmetry, the lattice point group and -- in the isotropic case -- the $SU(2)$ spin symmetry.
The intermediate phases PKS and CS break the translation symmetry and the point group symmetry but not the continuous $U(1)$ symmetry of the Hamiltonian.
The CS spin phase preserves the $60^\circ$ rotations of the triangular lattice around one site in the unit cell
while the PKS phase obeys $60^\circ$ rotations about a different site in the unit cell only in combination with a global flip of all spins.
Finally, the magnetic phase breaks the continuous $U(1)$ ($SU(2)$ for the isotropic case) spin symmetry down to a remaining $Z_2$ symmetry.

In the following, we present an in-depth analysis of the different phases starting with the disordered and magnetic phases in Sec.~\ref{subsec:disordered} and Sec.~\ref{subsec:magnetic}.
Afterwards, we discuss the intermediate region and its possible internal transitions in Sec.~\ref{subsec:pks_vs_cs} and demonstrate its exotic character induced by the geometrical frustration and
the strong competition between the PKS and CS ground state.
Finally, we present data for the quantum phase transitions from the intermediate region to the magnetic and disordered phase in Sec.~\ref{subsec:phase_transitions}.

\subsection{Disordered phase}
\label{subsec:disordered}
The disordered phase directly relates to the limit $J_K\to\infty$ which corresponds to the atomic limit.
In this limit, the unique ground state is the direct tensor product of on-site singlets formed by the spins of the two layers.
The rest of the many-body spectrum is separated by a large gap of order $J_K$.
Standard nondegenerate perturbation theory connects the atomic limit to finite $J_K$.
The first contribution is at order $J^2/J_K$.
The large gap in this phase leads to a fast exponential decay of correlations and allows for an exact description of the ground state as a tensor network with small bond dimension.
Furthermore, the ground state respects all symmetries of the Hamiltonian.

\subsection{Magnetic phase}
\label{subsec:magnetic}
\begin{figure}
\subfloat[Upper layer $I_z=0.4J$ \label{sfig:mag_upperlayer}]{%
  \includegraphics[width=0.4\linewidth]{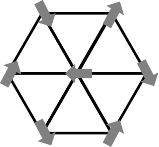}%
} \qquad
\subfloat[Lower layer $I_z=0.4J$ \label{sfig:mag_lowerlayer}]{%
  \includegraphics[width=0.4\linewidth]{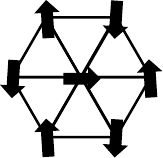}%
} \hfill
\subfloat[Upper layer $I_z=0$ \label{sfig:mag_upperlayer_Iz0}]{%
  \includegraphics[width=0.4\linewidth]{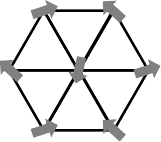}%
} \qquad
\subfloat[Lower layer $I_z=0$ \label{sfig:mag_lowerlayer_Iz0}]{%
  \includegraphics[width=0.4\linewidth]{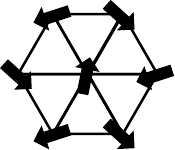}%
} \hfill
\caption{Magnetization patterns for both layers in the magnetic phase as obtained by iPEPS simulations with $D=5$ for $J_K=0.2J$.}
\label{fig:mag_patterns}
\end{figure}
\begin{figure}
\includegraphics[width=0.98\columnwidth]{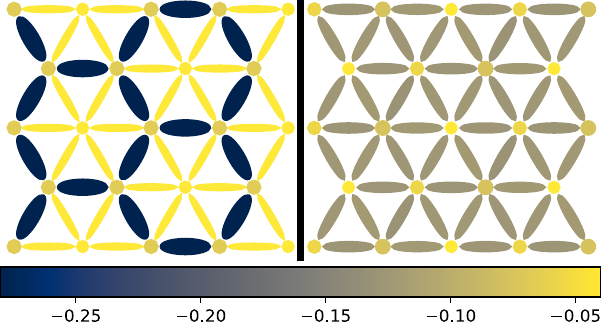}
\caption{
  Visualization of the ground state energy as obtained by iPEPS simulations with $D=5$ for $J_K=0.2J$. Left: clock-phase at $I_z=0.4J$, right: $120^\circ$-phase at $I_z=0$.
  Bond patches display the summation of the contributions from both layers: $J\langle\pmb{S}_i\pmb{S}_j\rangle + I_z\langle S^z_iS^z_j\rangle$.
  Site patches display the Kondo energy: $J_K\langle\pmb{S}_i\pmb{s}_i\rangle$.}
\label{fig:mag_energy}
\end{figure}
The magnetic phase is the second paradigmatic phase in the Doniach competition where the indirect magnetic interactions dominate.
It is realized for weak $J_K$. This section is partitioned into the case $I_z>0$ and $I_z=0$.

\subsubsection{Finite $I_z$}
The magnetization pattern for $I_z=0.4J$ and $J_K=0.2J$ is displayed in Figs.~\ref{fig:mag_patterns}\subref*{sfig:mag_upperlayer}--\subref*{sfig:mag_lowerlayer}.
Here, we show the expectation value $\langle\pmb{s}_i\rangle$ for the upper layer (Fig.~\ref{sfig:mag_upperlayer})
and the expectation value $\langle\pmb{S}_i\rangle$ for the lower layer (Fig.~\ref{sfig:mag_lowerlayer}).
Due to the weak interlayer coupling $J_K$, the two layers are noncollinear but in each layer, the pattern can be understood by looking at the decoupled layers.
The upper layer forms a spin-$1/2$ Heisenberg model which is known to have a symmetry broken ground state with three-sublattice $120^\circ$ order \cite{WC07}
and homogeneous spin correlations.
The absolute value of the magnetization is $m=\left|\langle\pmb{s}_i\rangle\right| \approx 0.33$ for the $D=5$ iPEPS simulations.
Although this value will decrease in the $D=\infty$ limit, it is substantially higher than the value of the triangular lattice Heisenberg model ($m\approx 0.2$~\cite{WC07}).
This is attributed to the residual influence of the Kondo interaction since we see a clear decrease of $m$ for decreasing $J_K$.

In contrast, the lower layer is an Ising model on the triangular lattice with Ising exchange $I_z$.
Its isolated ground state is strongly degenerate at finite $I_z$.
But a small transverse magnetic field lifts this degeneracy towards a clock ordered ground state~\cite{IM03}.
Since the main role of the transverse magnetic field in the Ising model is the creation of spin flips in the classical Ising state,
the finite magnetic moments of the upper layer act similar as a transverse magnetic field due to a weak $J_K$.
The only difference is that the effective field is inhomogeneous due to the three-sublattice $120^\circ$ order.
The spins in the lower layer are strongly polarized with a magnetization $M=\left|\langle\pmb{S}_i\rangle\right| \approx 0.49$, very close to the maximal value $M=1/2$.
The clock ordered ground state breaks the translational symmetry so that also the spin correlations along the nearest-neighbor bonds become inhomogeneous.
This can be seen in the left panel of Fig.~\ref{fig:mag_energy}, where the inhomogeneous bond energies are caused by the lower layer contribution $I_z\langle S_i^zS_j^z\rangle$.

\subsubsection{Vanishing $I_z$}

The magnetization pattern for $I_z=0$ is represented in Figs.~\ref{fig:mag_patterns}\subref*{sfig:mag_upperlayer_Iz0}--\subref*{sfig:mag_lowerlayer_Iz0}.
In this case, the two layers are collinear, i.e. spins on opposite sites are anti-parallel.
The upper layer has not changed qualitatively as compared to $I_z=0.4$. However in the decoupled limit, the lower layer is now completely degenerate.
Due to a weak $J_K$, the spins in the lower layer align antiparallel with their opposite partner so that the lower layer has the same $120^\circ$ order.
Therefore, the translational symmetry of the bond energies is preserved in the ground state which can be seen at the right of Fig.~\ref{fig:mag_energy}.
For the strength of polarization, we find similar values as compared to $I_z=0.4J$.

\begin{figure}
\includegraphics[width=0.98\columnwidth]{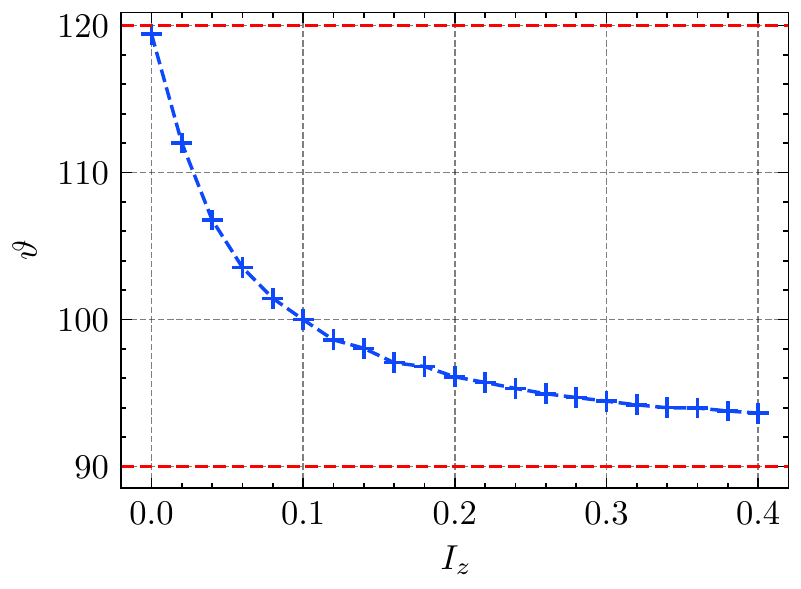}
\caption{The angle $\vartheta$ between adjacent spins as a function of $I_z$ as obtained by iPEPS simulations with $D=5$.
  Red dashed lines mark the two limiting cases $\theta=90^\circ$ for $I_z\to\infty$ and $\theta=120^\circ$ for $I_z=0$.}
\label{fig:angles}
\end{figure}

The clock order in the lower layer for finite $I_z$ is not exact but slightly canted due the finite coupling of the layers.
For $I_z=0.4J$, we find an angle of $\vartheta\approx 94^\circ$ instead of $\vartheta_{\text{clock}}=90^\circ$ for the perfect clock order.
The evolution of $\vartheta$ as a function of $I_z$ can be seen in Fig.~\ref{fig:angles}.
It shows that the angle evolves continuously from the $90^\circ$ clock order to the isotropic $120^\circ$ state.

\subsection{Intermediate phases}
\label{subsec:pks_vs_cs}
On a bipartite lattice, the disordered and the magnetic phases are separated by a quantum critical point where the competition is the strongest.
Adding geometrical frustration as a third competitor can lead to a region with new exotic phases in between the disordered and magnetic phase
where the Kondo screening is only partially active.
For the present triangular lattice, we find an intermediate phase \emph{only} at finite $I_z$ while the isotropic limit ($I_z=0$) shows the same behavior as in bipartite lattices.
The evidence for this is described in Sec.~\ref{subsec:phase_transitions} while this section focuses on the nature of the intermediate phase for finite $I_z$.

\begin{figure}[h]
\includegraphics[width=0.98\columnwidth]{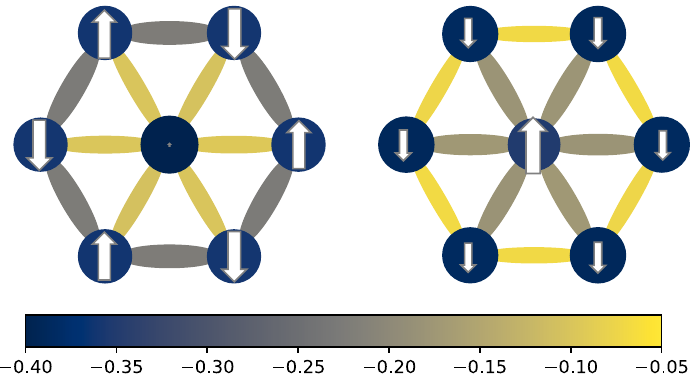}
\caption{The energy expectation values for the bond and site terms of the Hamiltonian and
  the onsite magnetization in the lower layer for the PKS phase (left) and CS phase (right) for an iPEPS with $D=6$.
  Parameters are $I_z=0.4J$, $J_K=0.7J$ and the bond contributions from the two layers are summed. Color legend refers to the bond energies.}
\label{fig:energy_pks_cs}
\end{figure}
We find \emph{two} different ground state candidates in competition with opposite character.
Their magnetization pattern for the lower layer and energy expectation values are displayed in Fig.~\ref{fig:energy_pks_cs}.
The magnetization pattern of the upper layer (not shown) is found to be the opposite of the lower layer.
The left state shows PKS where the central spin in a hexagon is completely screened.
The screening leads to a large onsite Kondo energy contribution for the central site.
The bond energies are higher along the circumference of the hexagon representing strong antiferromagnetic correlations between spins on the edge of the hexagon.
In contrast, the right panel shows a CS state
where the central site is not screened but strongly polarized leading to a low Kondo contribution of the center.
Reminiscent to a central spin model, the outer spins align antiparallel to the strongly polarized central spin while being partially screened.
In complete opposite to the PKS state, the bond energies are low along the circumference and high towards the center.
In that sense the CS state is an ``inverted'' version of the PKS state.

\begin{figure}
\subfloat[$x=0$ \label{sfig:mixing_CS}]{%
  \includegraphics[width=0.25\linewidth]{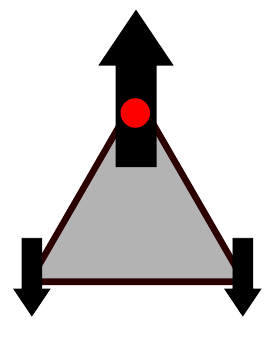}%
} \qquad
\subfloat[$x=1/2$ \label{sfig:mixing_MIX}]{%
  \includegraphics[width=0.25\linewidth]{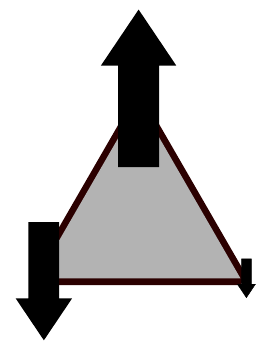}%
} \qquad
\subfloat[$x=1$ \label{sfig:mixing_PKS}]{%
  \includegraphics[width=0.25\linewidth]{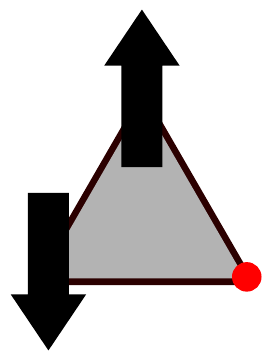}%
}
\caption{
  Illustration of the continuous mixing of both phases with three examples for the mixing degree $x$ (Eq.~\ref{eq:mixing}).
  The CS phase is realized for $x=0$ while the PKS phase corresponds to $x=1$.
  The red dot marks the central site of the hexagons in Fig.~\ref{fig:energy_pks_cs}.}
\label{fig:mixing}
\end{figure}

The two states in Fig.~\ref{fig:energy_pks_cs} can also be in an arbitrary superposition with the total energy being nearly independent of the superposition.
The degree of superposition can be parameterized by a single parameter utilizing the fact
that the total magnetization per layer $\sum_{i\in \text{cell}}\pmb{M}_i=\sum_{i\in \text{cell}}\pmb{m}_i=0$ vanishes.
The magnetization $M_i^z$ (with $i$ the nonequivalent sites in the unit cell) of the lower layer can then be parameterized with a single dimensionless parameter $x$,
\begin{align}
  \label{eq:mixing}
  M_1^z &= \frac{M(1+x)}{2}, \nonumber \\
  M_2^z &= \frac{M(1-x)}{2}, \\
  M_3^z &=  -M. \nonumber
\end{align}
Here, $M$ is the strength of the polarization of the moment pointing in opposite direction as the two other moments.
The PKS state corresponds to $x=1$ and the CS state to $x=0$. See Fig.~\ref{fig:mixing} for an illustration.

The strong competition of both ground state candidates makes the iPEPS simulations challenging.
E.g., the simple update simulations clearly favor the PKS ground state which indicates that an accurate consideration of the environment is required
to capture the CS state and the competition between both.
On account of this, we do not use the simple update at all but follow the subsequent protocol to identify the true ground state.
Simulations are initialized with random tensors at $D=3$. Afterwards the variational parameters are optimized using imaginary time evolution with the cluster update.
The result is taken as the initial state for the gradient-based optimization with AD.
To increase the bond dimension, we use again the imaginary time evolution in combination with the cluster update.
\begin{figure}
  \includegraphics{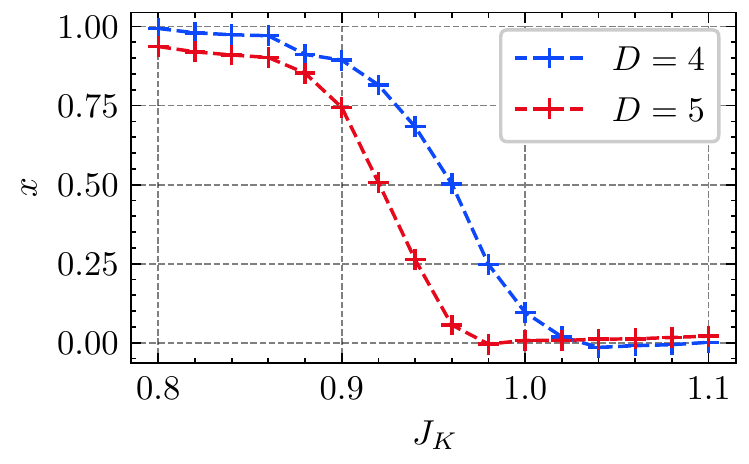}
  \caption{
    The mixture parameter $x$ (Eq.~\ref{eq:mixing}) as a function of $J_K$ for $I_z=0.4J$.
  }
\label{fig:mixing_vs_Jk}
\end{figure}

Figure~\ref{fig:mixing_vs_Jk} shows the mixture parameter $x$ as a function of $J_K$ for $I_z=0.4J$.
The simulations at $D=4$ and $D=5$ show a continuous change of $x$ from the CS state at stronger $J_K$ to PKS at weaker $J_K$.
Although the underlying tensors of the iPEPS are nonsymmetric in these cases, we do not see any $U(1)$ symmetry breaking in the observables.
This suggests that a description with symmetric tensors should be feasible.
However due to the strong competition between both phases, it turns out that it is very hard to find appropriate symmetry sectors in the tensors since these are determined at the initialization procedure
with an imaginary time evolution while the gradient based optimization works at fixed sectors. For an in-depth discussion of this issue see App.~\ref{sec:app}.

\begin{figure}
  \includegraphics{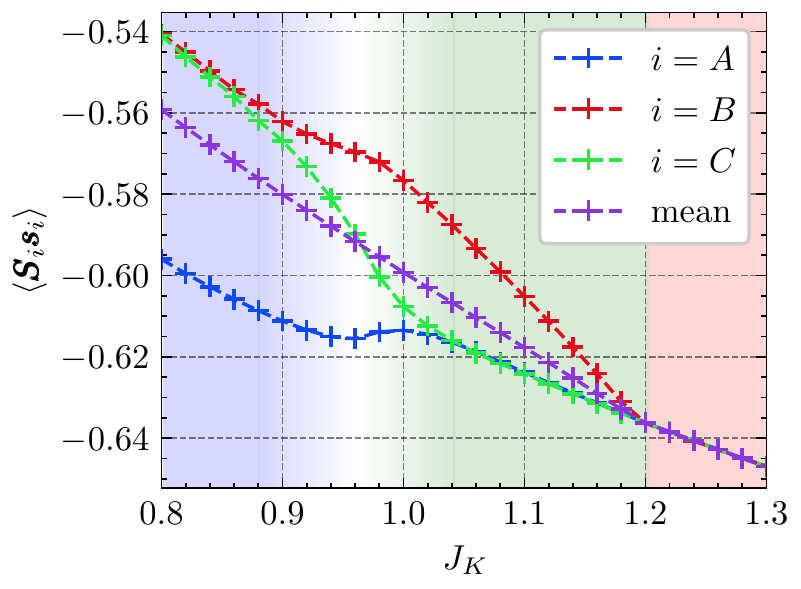}
  \caption{
    The site-resolved Kondo correlations for all three independent sites in the unit cell and the mean value for $I_z=0.4J$ at bond dimension $D=4$.
  }
\label{fig:site_kondo}
\end{figure}

In order to examine the physical properties of both phases, we analyze the site-resolved local Kondo correlations in Fig.~\ref{fig:site_kondo}.
For the chosen unit cell, there are three independent sites (see Fig.~\ref{sfig:ipeps_unitcell}) which are labeled as $A$, $B$ and $C$.
For $J_K>1.2J$, all three sites are equivalent and the system is in the disordered phase.
In the CS state (green shaded area in Fig.~\ref{fig:site_kondo}), one site is \emph{less} screened than the other two.
This site corresponds to the central site in the right hexagon of Fig.~\ref{fig:energy_pks_cs}.
On the other hand in the PKS phase (blue shaded area in Fig.~\ref{fig:site_kondo}), one site is \emph{more} screened than the other two.
Similarly, this site corresponds to the central site in the left hexagon of Fig.~\ref{fig:energy_pks_cs}.
In between, there is a broad transition regime in which all three sites have different local Kondo correlations.
Within the transition regime, one site is changing its nature from strongly to weakly screened. 

Regarding the nature of the transition, two different scenarios are conceivable:
(i) a first order direct transition between PKS an CS containing a jump in one of the local Kondo correlations and
(ii) a mixed phase in between PKS and CS including two continuous phase transitions.
This can also be corroborated by looking at the symmetries of the involved phases.
For the CS phase, the ground state obeys a $60^\circ$ rotational symmetry about the $B$-site and a $120^\circ$ rotational symmetry about the $A$- and $C$-site.
On the other hand, the PKS phase obeys a $60^\circ$ rotational symmetry in combination with a spin flip about the $A$-site and a $120^\circ$ rotational symmetry about the $B$- and $C$-site.
We note that the absolute position of the 6-fold axis is different in the PKS and CS phase and that in the PKS phase, an additional spin flip is required.
Thus, both phases have different symmetries which implies that a (conventional) direct transition would be of first order.
In the mixed phase, both 6-fold rotational symmetries are broken which allows two continuous transitions, first from CS to the mixed phase and then from the mixed phase to PKS.

Our iPEPS simulations do not show a direct transition between the two phases, and the size of the transition regime is almost equal for $D=4$ and $D=5$ (see Fig.~\ref{fig:mixing_vs_Jk}). This suggests that the appearance of the transition regime is not a finite-$D$ effect, but rather that scenario (ii) is realized, i.e. that the transition regime corresponds to a mixed phase. However, for a clarification, higher bond dimensions would be required.

The observed site-resolved Kondo correlations suggest a simple qualitative picture.
The multiplicity of the undermost line in Fig.~\ref{fig:site_kondo} changes from three in the disordered to two in the CS state and to one in the PKS state.
This can be interpreted as a gradual decrease of the Kondo screening with decreasing $J_K$ in a step-like manner.
However, this picture does not show up in the mean value of the local Kondo correlations which is featureless (see Fig.~\ref{fig:site_kondo}).
\subsection{Transitions}
\label{subsec:phase_transitions}
\subsubsection{Intermediate to disordered}
\begin{figure}
  \includegraphics{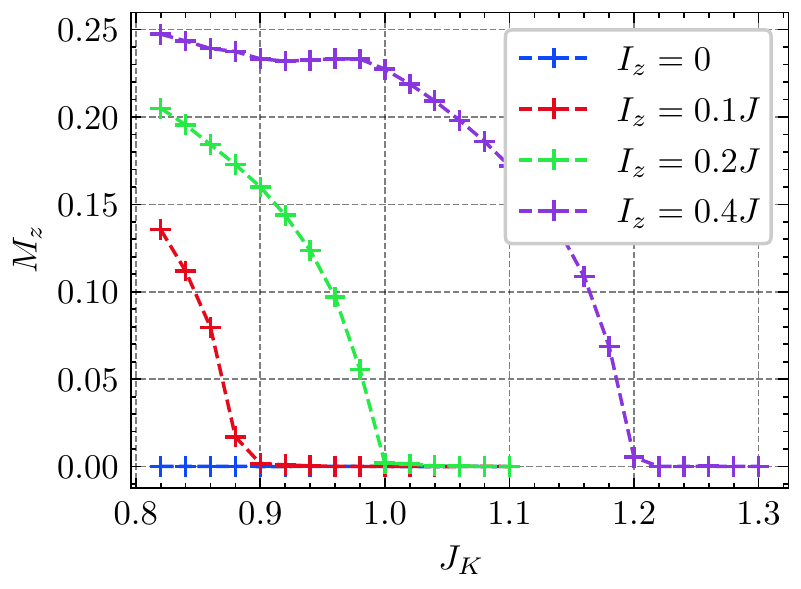}
  \caption{
    The order parameter $M_z$ as a function of $J_K$ as obtained by iPEPS simulations with bond dimension $D=4$.
  }
\label{fig:transition_d_i_Iz}
\end{figure}
\begin{figure}
  \includegraphics{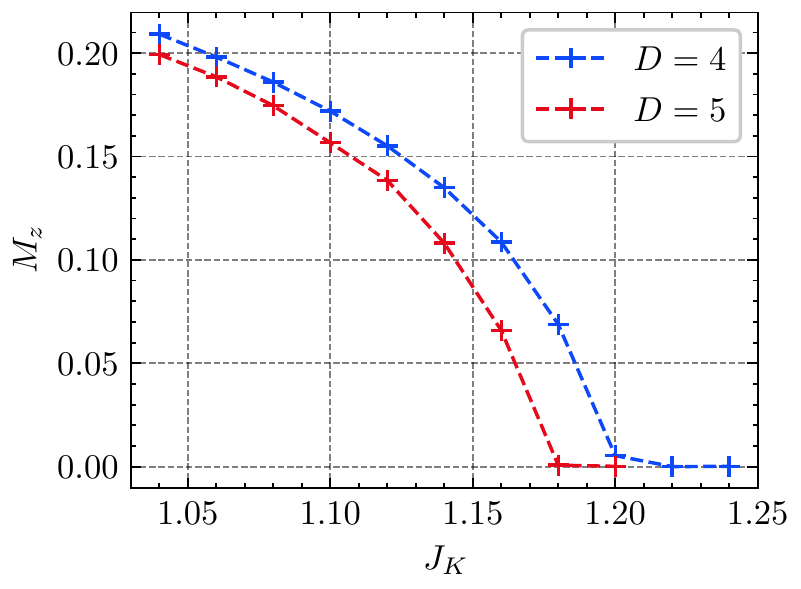}
  \caption{
    The order parameter $M_z$ as a function of $J_K$ for $I_z=0.4J$.
  }
\label{fig:transition_d_i}
\end{figure}

The disordered phase at strong $J_K$ consists of localized on-site Kondo singlets.
The decrease of $J_K$ gradually increases the nonlocal correlations and the entanglement in the ground state
and can eventually lead to a phase transition.
The phase transition goes along with a symmetry breaking of the SU(2) spin symmetry so that the spontaneous magnetization $M_z=\frac{1}{N}\sum_{i\in\text{cell}}\left| \langle S_i^z\rangle\right|$ serves as an order parameter. 
To detect this transition point, we study the magnetization $M_z$ in the lower layer as a function of $J_K$.
Figure~\ref{fig:transition_d_i} shows the iPEPS results of the order parameter at bond dimension $D=4$ for several values of $I_z$.
The data is consistent with a continuous quantum phase transition which shifts to larger $J_K$ for increasing $I_z$ in agreement with Ref.~\onlinecite{MNY10}.
Beside this qualitative agreement, the quantitative transitions points are lower than the values found in Ref.~\onlinecite{MNY10}.
E.g. for $I_z=0.2J$, we find $J_K^c\approx J$ as compared to $J_K^c\approx 1.2J$.
In big contrast to the study in Ref.~\onlinecite{MNY10}, we do not encounter a transition to the intermediate phase in the isotropic case $I_z=0$.
Instead, we find a direct transition to the magnetic phase as discussed in Sec.~\ref{subsubsec:mag_to_inter}.

To check the $D$-dependence, we plot the transition for $I_z=0.4J$ for $D=4$ and $D=5$.
This shows that the transition point shifts to smaller values of $J_K$, i.e. shifts further away from the behavior found in Ref.~\onlinecite{MNY10}.
\subsubsection{Magnetic to intermediate}
\label{subsubsec:mag_to_inter}
\begin{figure}
  \includegraphics{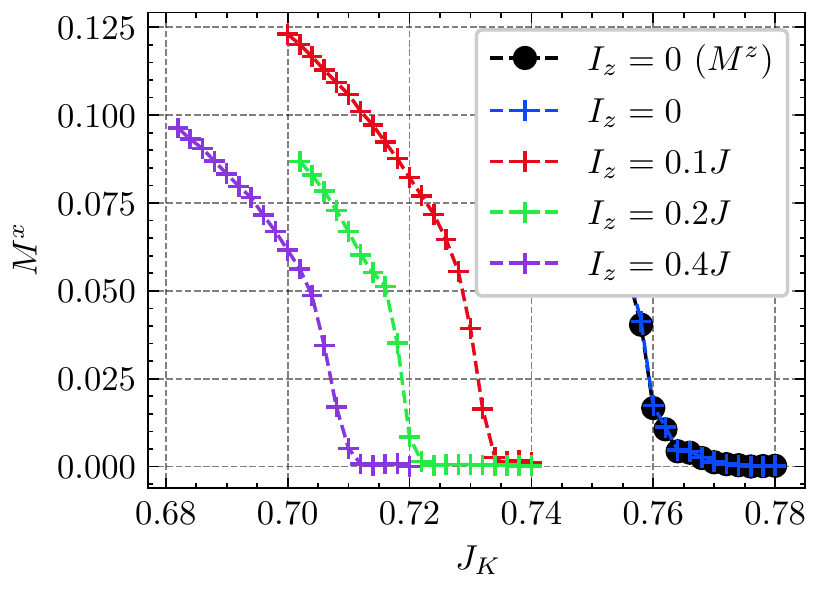}
  \caption{
    The order parameter $M_x$ (except for the black circles) as a function of $J_K$ for several values of $I_z$ with iPEPS bond dimension $D=4$.
    The transition point $J_K^{i\leftrightarrow m}$ is only weakly dependent on $I_z$.
  }
\label{fig:transition_i_m}
\end{figure}
The second phase transition is connected with the breaking of the remaining U(1) spin symmetry and can be observed by the order parameter $M_x=\frac{1}{N}\sum_{i\in\text{cell}}\left| \langle S_i^x\rangle\right|$.
We plot the results of our iPEPS simulations in Fig.~\ref{fig:transition_i_m} for several values of $I_z$.
The transition point is only weakly dependent on $I_z$ and we find $J_K^{i\leftrightarrow m}\approx 0.7J$.
This is approximately twice as large as the value found in Ref.~\onlinecite{MNY10}.
As opposed to Ref.~\onlinecite{MNY10}, the transition is not completely independent of $I_z$ and we find that the transition point shifts to larger values of $J_K$ for weaker $I_z$.
The data for the order parameter is consistent with a continuous quantum phase transition.

For $I_z=0$, the transition is different because the intermediate phase is missing,
i.e. at the right of Fig.~\ref{fig:transition_i_m}, the ground state is disordered.
This can be seen by comparing the value of $M_x$ with that of $M_z$ (black circles in Fig.~\ref{fig:transition_i_m}).
Both are zero above the phase transition and increases equally indicating a homogeneous $120^\circ$ order.
\section{Conclusions and outlook}
\label{sec:con}
We have studied the influence of geometrical frustration on the two-dimensional Kondo-necklace model using iPEPS.
Our findings include two novelties as compared to previous studies:
(i) the PKS ground state is \emph{not} realized in the isotropic Kondo-necklace model.
I.e. this exotic state is clearly driven by anisotropic interactions and not only by geometrical frustration.
This can be compared to the simplest model for quantum magnetism, the Heisenberg model.
The unfrustrated square lattice is magnetically ordered.
This order persists for the geometrically frustrated triangular lattice however with stronger quantum fluctuations~\cite{WC07}.
For even more frustrated lattices, e.g. the Kagomé lattice, the quantum fluctuations destroy the magnetic order leading to a spin liquid ground state~\cite{SHW11, DMS12, IBS13, MCH17, LXC17, HZO17}.
This comparison raises the question if the PKS or any other exotic intermediate phase will be present in the Kondo-necklace model on stronger frustrated lattices.
(ii) The PKS ground state is \emph{not} the only exotic phase driven by the geometrical frustration in combination with anisotropic interactions.
The second candidate has a strongly polarized spin in the center of a hexagon pointing antiparallel to its surrounding weakly polarized neighbors.
We named it central spin (CS) phase and it can be seen as an inverted version of the PKS phase.
We find that both phases are in an extraordinary strong competition which makes the iPEPS simulations challenging.
Both a direct transition and a mixed phase in between PKS and CS are conceivable, but our iPEPS simulations 
are more compatible with the latter. 

Moreover, we have unveiled the  $J_K-I_z$ phase diagram and found quantitative differences to previous studies for all phase boundaries.
At weak $J_K$, the ground state is magnetically ordered and at strong $J_K$ the ground state is disordered.
The magnetic order at weak $J_K$ is of clock-type for strong $I_z$ and goes smoothly into a $120^\circ$ order for $I_z\to 0$.
The magnetic and disordered phase touch each other directly for $I_z=0$ while there are two intermediate phases in between at finite $I_z$.
The first one after the disordered is PKS which is in strong competition with the second, the previously mentioned new CS spin phase. 
The order of the phases can be understood qualitatively by the simple rule that the amount of Kondo screening is decreasing with decreasing $J_K$.
In the disordered phase, three out of three sites are screened. In the CS spin phase this reduces to two out of three and in the PKS phase only one out of three sites is screened.
Finally in the magnetically ordered phase, no Kondo screening is present.

Our work is potentially relevant for recently discovered van der Waals heterostructures~\cite{VAG21}.
These structures constitute very clean two-dimensional systems on the triangular lattice where the microscopic parameters can be controlled by the twist of the heterostructures.
Especially, also the filling can be tuned via gating the samples.
Thus, it is conceivable that similar physics as in our case can be studied when probing their magnetic properties.

Two different routes can be chosen to further explore the influence of geometrical frustration to Kondo lattice systems.
First, one can increase the geometrical frustration by considering more frustrated lattices or by including further interactions.
Second, one can fully include the electronic layer and study the full Kondo lattice model.
This is in particular interesting since the electronic layer can mediate diverse effective interactions and additionally can be doped away from half-filling.
This would shrink the gap to the experimental realizations of frustrated Kondo lattice systems since the main compounds CePdAl and $\text{UNi}_4\text{B}$ are heavy-fermion metals.
\section*{Acknowledgment}
We are grateful to Yukitoshi Motome for useful discussions.

M.P. is funded by the Deutsche Forschungsgemeinschaft (DFG, German Research Foundation) -- project ID 497779765.

This project has received funding from the European Research Council (ERC) under the European Union's Horizon 2020 research and innovation programme (grant agreement Nos. 677061 and 101001604).

\bibliography{lit,lit_new,litplus,refs}
\clearpage
\appendix
\begin{widetext}
\section{Symmetric tensors in the intermediate phase}
\label{sec:app}
The intermediate coupling regime includes the competition between the PKS and CS ground state.
In our simulations with nonsymmetric tensors, we do not see a $U(1)$ symmetry breaking in the observables.
This suggests that a description with symmetric tensors should be feasible.
In practical computations however, it turns out that the results scatter between both possible ground states.

\begin{figure}
  \includegraphics{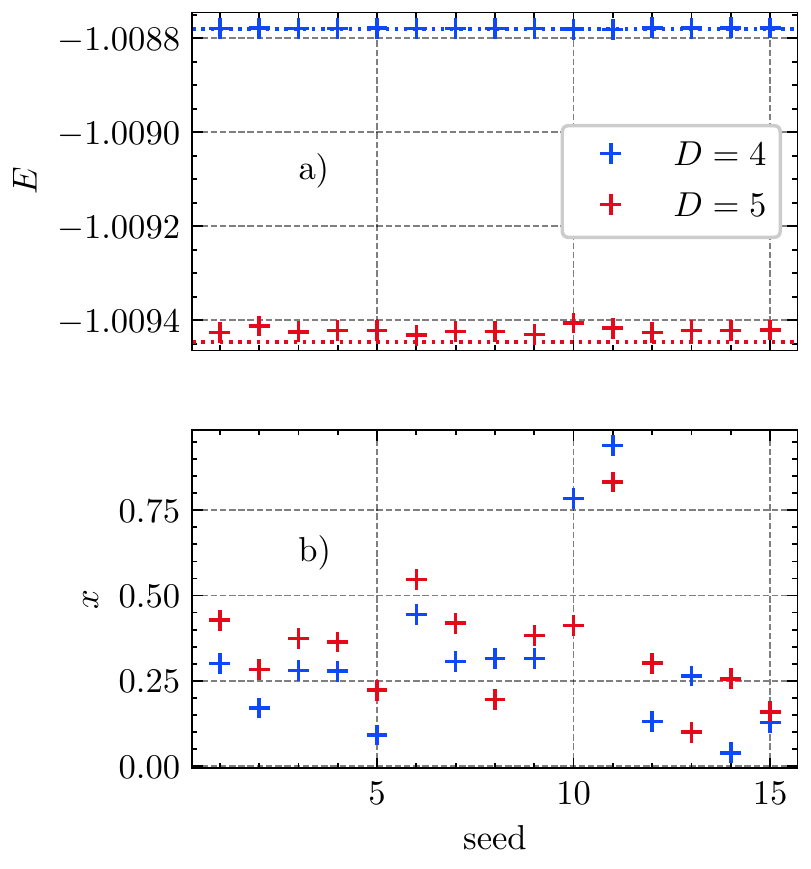}
  \caption{
    a) Variational energies for various different random initial sector distributions at $J_K=J$ and $I_z=0.4J$. The dotted line marks the energy with nonsymmetric tensors in the iPEPS ansatz.
    b) The mixture parameter $x$ for the various seed values.
  }
\label{fig:symmetric_jk1}
\end{figure}
To demonstrate this issue we focus on $J_K=J$ and run simulations with various random sector distributions.
The simulations are initialized with random charges for each nonequivalent bond in the iPEPS tensor network at bond dimension $D=3$.
Thereafter, imaginary time evolution with the cluster update is used to optimize the Ansatz.
We note that during this process the charge sectors are allowed to change.
After increasing the bond dimension to $D=4$, we run gradient based optimization with AD and take the simulations with lowest energy.
In that case, these are the results of 15 different simulations.
With an additional imaginary time evolution -- which may change the charge sectors again -- the bond dimension is increased to $D=5$ followed by another gradient based simulation.
The energies (mixture parameters) of the different simulations are displayed in the top (bottom) of Fig.~\ref{fig:symmetric_jk1}. 
The mixture parameters scatter over nearly the whole range $[0,1]$ with a cumulation at intermediate values.
In contrast, the final energies lie very close. Compared to the energy obtained with nonsymmetric tensors (dotted lines in Fig.~\ref{fig:symmetric_jk1}), all obtained energies are slightly higher.
However, the relative distance is of order $10^{-5}$.

\begin{figure}
  \includegraphics{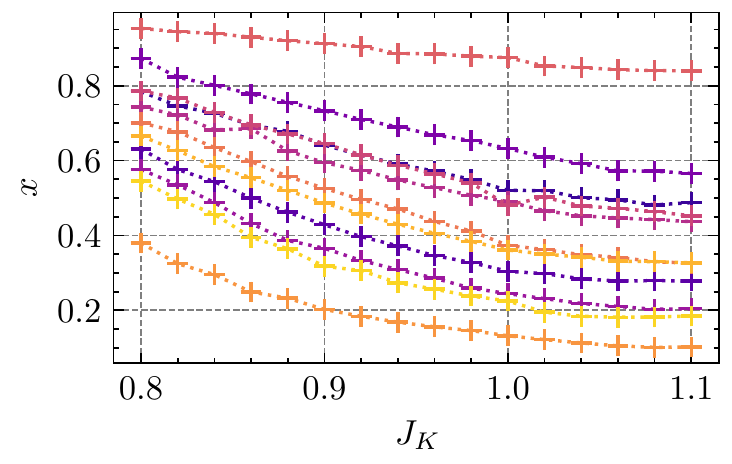}
  \caption{
    The mixing parameter as function of $J_K$ for $I_z=0.4J$ and various seed values at bond dimension $D=5$.
  }
\label{fig:symmetric_mixing}
\end{figure}
Finally, the results at $J_K=J$ are used as initial states for the whole intermediate $J_K$ range as displayed in Fig.~\ref{fig:symmetric_mixing}.
These initial states are optimized using gradient based optimization with AD which does not change the charge sectors.
The overall tendency that the mixture increases with decreasing $J_K$ is in agreement with the nonsymmetric simulations.
\end{widetext}
\end{document}